\documentclass[11pt,showpacs,showkeys,aps,pra]{revtex4}
\usepackage[utf8x]{inputenc}
\usepackage{ucs}

\usepackage{amsmath,amssymb}
\usepackage{latexsym}
\usepackage{amsfonts}
\usepackage{bm}
\usepackage[usenames]{color}
\usepackage{multirow}
\usepackage{graphicx}
\usepackage{hyperref}
\usepackage{subfig}
\usepackage{xcolor}
\usepackage{tabularx}
\usepackage[normalem]{ulem}
\usepackage{float} 
\usepackage{wrapfig} 
\usepackage{upgreek} 
\usepackage{cancel} 
\usepackage{mathdots} 
\usepackage{mathrsfs} 
\usepackage{mathtools}
\usepackage{threeparttable}
\usepackage{standalone}
\usepackage{booktabs, dcolumn}
         
\AtBeginDocument{%
  }

\graphicspath{{figures/}} 

\begin{document}

\title{Rényi, Shannon and Tsallis entropies of Rydberg hydrogenic systems}

\author{I.V. Toranzo}
\email[]{ivtoranzo@ugr.es}
\affiliation{Departamento de F\'{\i}sica At\'{o}mica, Molecular y Nuclear and Instituto Carlos I de F\'{\i}sica Te\'orica y Computacional, Universidad de Granada, Granada 18071, Spain}

\author{J.S. Dehesa}
\email[]{dehesa@ugr.es}
\affiliation{Departamento de F\'{\i}sica At\'{o}mica, Molecular y Nuclear and Instituto Carlos I de F\'{\i}sica Te\'orica y Computacional, Universidad de Granada, Granada 18071, Spain}
\begin{abstract}
The Rényi entropies $R_{p}[\rho], 0<p<\infty$ of the probability density $\rho_{n,l,m}(\vec{r})$ of a physical system completely characterize the chemical and physical properties of the quantum state described by the three integer quantum numbers $(n,l,m)$. The analytical determination of these quantities is practically impossible up until now, even for the very few systems where their Schr\"odinger equation is exactly solved. In this work, the  Rényi entropies of Rydberg (highly-excited) hydrogenic states are explicitly calculated in terms of the quantum numbers and the parameter $p$. To do that we use a methodology which first connects these quantities to the $\mathcal{L}_{p}$-norms $N_{n,l}(p)$ of the Laguerre polynomials which characterize the state's wavefunction. Then, the Rényi, Shannon and Tsallis entropies of the Rydberg states are determined by calculating the asymptotics ($n\rightarrow\infty$) of these Laguerre norms. Finally, these quantities are numerically examined in terms of the quantum numbers and the nuclear charge.
\end{abstract}

\pacs{89.70.Cf, 89.70.-a, 32.80.Ee, 31.15.-p}

\keywords{Information theory of quantum systems, Rényi entropy of quantum systems, Rydberg states, hydrogenic atoms.}

\maketitle

\section{Introduction}

Recent years have witnessed a growing interest in the analytical information theory of finite quantum systems. A major goal of this theory is the explicit determination of the entropic measures (Fisher information and Shannon, Rényi and Tsallis entropies,...) in terms of the quantum numbers which characterize the state’s wavefunction of the system. These quantities, which quantify the spatial delocalization of the single-particle density of the systems in various complementary ways, are most appropriate uncertainty measures because they do not make any reference to some specific point of the corresponding Hilbert space, in contrast to the variance and other dispersion measures. Moreover, they are closely related to numerous energetic and experimentally measurable quantities of the system \cite{gadre_1987,parr-yang,liu-parr1,liu-parr2,nagy,gadre_2003,jizba_2004,dehesa_sen12} and they have been used as indicators of various atomic and molecular phenomena \cite{rosario1,rosario2,rosario3,calixto,He-2015,Mukherjee}. Since the Schrödinger equation can be exactly solved only for a few quatum-mechanical potentials which model most of the quantum chemical and physical phenomena, most of the efforts have been focused on the harmonic and hydrogenic systems up until now. Basically this is because the wavefunctions of their ground and excited states are controlled by the hypergeometric orthogonal polynomials (Hermite, Laguerre, Jacobi) whose analytical properties are under control.\\

 Apart from the Fisher information whose explicit values have been found \cite{romera_2005,dehesa_2006}, the entropic measures of the oscillator-like and hydrogenic systems have not yet been analytically determined for all quantum-mechanically-allowed states except for the ground and a few low-lying states \cite{gadre_1985,yanez_1994,bhattacharya,ghosh,dehesa_2010,lopez_2013} and for the high-lying states in the Shannon case (see Eq. (16) in \cite{lopez_2013}). In addition some rigorous bounds on these entropic measures, as well as some related uncertainty relations, have been found \cite{sanchezmoreno,guerrero,rudnicki,jizba}.\\

The Rényi entropies $R_{p}[\rho]$ and Tsallis entropies $T_{p}[\rho]$ of a probability density $\rho(\vec{r})$ are defined \cite{renyi1,tsallis} as
\begin{equation}
\label{eq:renentrop}
R_{p}[\rho] =  \frac{1}{1-p}\ln W_{p}[\rho]; \quad 0<p<\infty,
\end{equation}
\begin{equation}
\label{eq:renentrop}
T_{p}[\rho] =  \frac{1}{p-1}(1-W_{p}[\rho]); \quad 0<p<\infty,
\end{equation}
where $W_{p}[\rho]$ denotes the $p$-th entropic moment of $\rho(\vec{r})$ which is given by 
\begin{equation}
\label{eq:entropmom}
W_{p}[\rho] = \int_{\mathbb{R}^3} [\rho(\vec{r})]^{p}\, d\vec{r}, \quad p\ge 0, 
\end{equation}
These quantities completely characterize the density $\rho(\vec{r})$ \cite{romera_01,jizba_2015b}.
Note that these quantities include the Shannon entropy (since $S[\rho] := \int \rho(\vec{r}) \ln \rho(\vec{r}) d\vec{r}  = \lim_{p\rightarrow 1} R_{p}[\rho] = \lim_{p\rightarrow 1} T_{p}[\rho]$), and the disequilibrium, $\langle\rho\rangle = \exp(R_{2}[\rho])$, as two important particular cases. Moreover, they are mutually connected by the relation
\begin{eqnarray}
\label{eq:rel2}
T_{p}[\rho] &=& \frac{1}{1-p}[e^{(1-p)R_{p}[\rho]}-1] ,
\end{eqnarray}
For a revision of the Rényi entropies properties see \cite{aczel,dehesa_89,romera_01,leonenko,guerrero,jizba} and the reviews \cite{dehesa_sen12,bialynicki3,jizba}. The R\'enyi entropies and their associated uncertainty relations have been widely used to investigate a great deal of quantum-mechanical properties and phenomena of physical systems and processes \cite{bialynicki2,dehesa_sen12,bialynicki3}, ranging from the quantum-classical correspondence \cite{sanchezmoreno} and quantum entanglement \cite{bovino} to pattern formation and Brown processes \cite{cybulski1,cybulski2}, quantum phase transition \cite{calixto}, disordered systems \cite{varga} and multifractal thermodynamics \cite{jizba_2004b}. Moreover, there exist various classical and quantum coding theorems \cite{campbell,jizbak} which endow the Rényi and Tsallis entropies with an operational (so, experimentally verifiable) meaning. \\

In this work we analytically determine the Shannon, Rényi and Tsallis entropies of the highly-excited (Rydberg) hydrogenic states on the same footing, by use of a methodology based on the strong asymptotics of Laguerre polynomials. The Rydberg states \cite{gallagher,lundee} play a relevant role from both fundamental and applicable points of view. Indeed they can be considered a fertile laboratory where to investigate the order-to-chaos transitions through the applications of electric fields and, because of their extraordinary properties, they are being presently used in many technological areas such as e.g. quantum information processing \cite{shiell,saffman}. The entropic moments and their associated Shannon, Rényi and Tsallis entropies quantify the internal disorder of the Rydberg atom as given by its quantum probability density in a complementary, but much more complete, way than the variance and other dispersion measures whose values have been already shown \cite{dehesa_2010,lopez_2013}.\\

The structure of this work is the following. First, in sec. \ref{sec:2}, we state the problem and the methodology to solve it. In section \ref{sec:3} we obtain the radial Rényi entropy $R_{p}[\rho_{n,l}]$ of the Rydberg hydrogenic states for all possible values of the involved parameters in an analytical way. Then in section \ref{sec:4}, we obtain the final results of the Rényi, Tsallis and Shannon entropies for the Rydberg hydrogenic states and, in addition, we numerically compute the Rényi entropies for some specific Rydberg states and its variation with the nuclear charge of the atom. In section \ref{sec:5}, some conclusions are given.

\section{Statement of the problem}\label{sec:2}

Let us now determine the Rényi entropies of the Rydberg hydrogenic states characterized by the Coulombian potential $V_{D}(r) = - \frac{Z}{r}$. It is well-known that these states are given (see e.g., \cite{lopez_2013} and references therein) by the wavefunctions characterized by the energies $E_{n,l}= -\frac{Z^{2}}{2n^{2}}$ and the quantum probability densities
\begin{equation}
\label{eq:qpd}
\rho_{n,l,m}(\vec{r}) = \frac{4Z^{3}}{n^{4}}\frac{\omega_{2l+1}(\tilde{r})}{\tilde{r}}[\widehat{L}_{n-l-1}^{(2l+1)}(\tilde{r})]^{2}\,\,|Y_{l,m}(\theta,\phi)|^{2} \equiv \rho_{n,l}(\tilde{r})\,|Y_{l,m}(\theta,\phi)|^{2},
\end{equation}
where $\tilde{r}=\frac{2 Z}{n}r $, $n=1,2,3,\ldots$, $l = 0, 1,\ldots, n − 1$, $m= −l, −l + 1, \ldots, +l$, $\widehat{L}_{n}^{\alpha}(x)$ denotes the orthonormal Laguerre polynomials \cite{abramowitz} with respect to the weight function $\omega_{\alpha}=x^{\alpha}e^{-x}$ on the interval $[0,\infty)$, and $Y_{l,m}(\theta,\phi)$ denotes the spherical harmonics \cite{abramowitz} given by
\begin{equation}
\label{eq:sphharm}
Y_{l,m}(\theta,\phi) = \left(\frac{(l+\frac{1}{2})(l-|m|)![\Gamma(|m|+\frac{1}{2})]^{2}}{2^{1-2|m|}\pi^{2}(l+|m|)!} \right)^{\frac{1}{2}}e^{im\phi}(\sin\theta)^{|m|}C_{l-|m|}^{|m|+\frac{1}{2}}(\cos\theta).
\end{equation}
Then, by keeping in mind Eqs. (\ref{eq:renentrop})-(\ref{eq:entropmom}), the entropic moments of the hydrogenic state $(n,l,m)$ are
\begin{equation}
\label{eq:entropmom2}
W_{p}[\rho_{n,l,m}] = \int_{\mathbb{R}^3} [\rho_{n,l,m}(\vec{r})]^{p}\, d\,\vec{r}= \int\limits_{0}^{\infty}[\rho_{n,l}(r)]^{p}\,r^{2}\,dr \times \Omega_{l,m}(\theta,\phi),
\end{equation}
where the angular part
\begin{equation}
\label{eq:angpart}
\Omega_{l,m}(\theta,\phi) = \int_{0}^{\pi} \int_{0}^{2\pi}[Y_{l,m}(\theta,\phi)]^{2p}\sin\theta\, d\theta d\phi,
\end{equation}
and the Rényi entropies of the hydrogenic state $(n,l,m)$ can be expressed as
\begin{equation}
\label{eq:totalentro}
R_{p}[\rho_{n,l,m}] = R_p[\rho_{n,l}] + R_{p}[Y_{l,m}],
\end{equation}
where $R_p[\rho_{n,l}]$ denotes the radial part
\begin{equation}
\label{eq:radentropy}
R_p[\rho_{n,l}] = \frac{1}{1-p}\ln\int\limits_{0}^{\infty}[\rho_{n,l}(r)]^{p}\,r^{2}dr,
\end{equation}
and $R_{l,m}[Y_{l,m}]$ denotes the angular part
\begin{equation}
\label{eq:angentro}
R_{p}[Y_{l,m}] = \frac{1}{1-p} \ln \Omega_{l,m}(\theta,\phi),
\end{equation}
which is the Rényi-entropic functional of the well-controlled spherical harmonics \cite{abramowitz}.
Since the radial part is the only component which depends on the principal quantum number $n$, the crucial problem for the calculation of the Rényi entropy $R_{p}[\rho_{n,l,m}]$ for the Rydberg states (i.e., states with a very large $n$) of hydrogenic systems is to determine the value of the radial Rényi entropy $R_p[\rho_{n,l}]$ in the limiting case $n\to\infty$.
Taking into account (\ref{eq:radentropy}), the explicit expression of $\rho_{n,l}(\tilde{r})$ given by $(\ref{eq:qpd})$, and that the $\mathcal{L}_{p}$-norm of the Laguerre polynomials $\widehat{L}_{n}^{(\alpha)}(x)$ is
\begin{equation}\label{eq:c1.2}
N_{n}(\alpha,p,\beta)=\int\limits_{0}^{\infty}\left(\left[\widehat{L}_{n}^{(\alpha)}(x)\right]^{2}\,w_{\alpha}(x)\right)^{p}\,x^{\beta}\,dx,
\end{equation}
(with $\alpha > -1, p > 0$ and, to guarantee convergence at zero, $\beta+p\alpha > -1$), one has that the radial Rényi entropy can be expressed as
\begin{equation}
\label{eq:renyi_lp}
R_{p}[\rho_{n,l}] =\frac{1}{1-p}\, \ln\left[\frac{n^{3-4p}}{2^{3-2p}Z^{3(1-p)}}N_{n,l}(\alpha,p,\beta)\right],
\end{equation}
where the norm $N_{n,l}(\alpha,p,\beta) \equiv N_{n,l}( p)$ is given by
\begin{equation}\label{1.2}
N_{n,l}(\alpha,p,\beta) =\int\limits_{0}^{\infty}\left(\left[\widehat{L}_{n-l-1}^{(\alpha)}(x)\right]^{2}\,w_{\alpha}(x)\right)^{p}\,x^{\beta}\,dx,
\end{equation}
with 
\begin{equation}\label{eq:c1.3}
\alpha=2l+1\,,\;l=0,1,2,\ldots,n-1, \quad p>0\quad\mbox{and}\quad \beta=2-p\;.
\end{equation}
Note that (\ref{eq:c1.3}) guarantees the convergence of integral (\ref{1.2}) since the condition $\beta+p\alpha= 2(1+lp) > -1$  is always satisfied for the physically meaningful values of the parameters.\\

Thus, by keeping in mind Eqs. (\ref{eq:totalentro}), (\ref{eq:radentropy}), (\ref{eq:angentro})  and (\ref{eq:renyi_lp}), the determination of the Rényi entropy $R_{p}[\rho_{n,l,m}]$ for the Rydberg states entails the calculation of the asymptotics ($n\to\infty$) of the Laguerre norms $N_{n,l}(\alpha,p,\beta)$ given by Eqs. (\ref{1.2}) and (\ref{eq:c1.3}), which will be solved in the next section.

\section{Radial Rényi entropy $R_{p}[\rho_{n,l}]$ of Rydberg states}
\label{sec:3}

Let us here determine the radial entropy of the Rydberg hydrogenic states, i.e. the asymptotics ($n\to\infty$) of the radial Rényi entropy $R_{p}[\rho_{n,l}]$ which, according to Eq. (\ref{eq:renyi_lp}), essentially reduces to the asymptotics ($n\to\infty$) of the Laguerre norms $N_{n,l}(\alpha,p,\beta)$ given by (\ref{1.2}) and (\ref{eq:c1.3}). \\

To do that we use the method of Aptekarev et al which has been recently applied to oscillator-like systems \cite{aptekarev_2015}. This method allows us to find the asymptotics of the Laguerre functionals $N_{n}(\alpha,p,\beta)$ given by (\ref{eq:c1.2}) with $\alpha > -1, p > 0$ and $\beta+p\alpha > -1$. It shows that the dominant contribution in the magnitude of the integral (\ref{eq:c1.2}) comes from different regions of integration in~(\ref{eq:c1.2}); these regions depend on the different values of the involved parameters ($\alpha,p,\beta$). This entails that we have to use various asymptotical representations for the Laguerre polynomials at different regions of the interval of orthogonality ($0,\infty$).\\

Altogether there are five asymptotical regimes which can give (depending on $\alpha,\beta$ and $p$) the dominant contribution in the
asymptotics of $N_{n}(\alpha, p, \beta)$. In three of them (which we call by Bessel, Airy and cosine regimes) the involved Laguerre norm $N_{n}(\alpha, p, \beta)$ grows according to a power law in $n$ with an exponent which depends on $\alpha,\beta$ and $p$. The Bessel regime corresponds to the neighborhood of zero (i.e., at the left extreme of the orthogonality interval), where the Laguerre polynomials can be asymptotically described by means of Bessel functions (taken for expanding scale of the variable). Then (to the right of zero) the oscillatory behavior of the polynomials (in the bulk region of zeros location) is asymptotically modelled by means of the trigonometric functions (cosine regime) and at the neighborhood of the extreme right zeros asymptotics is given by Airy functions (Airy regime). Finally, at the extreme right of the orthogonality interval (i.e., near infinity) the polynomials have  growing asymptotics. Moreover, there are two transition regions (to be called by cosine-Bessel and cosine-Airy) where these asymptotics match each other; that is, asymptotics of the Bessel functions for big arguments match the trigonometric function, as well as the asymptotics of the Airy functions do the same. \\

The $n$th-power laws in the Bessel, Airy and cosine regimes are controlled by the constants $C_{B}(\alpha,p,\beta)$, $C_{A}(p)$ and $C(\beta,p)$, respectively, whose values are given in Table \ref{table1}. Therein, we have used the notation
$$
J_{\alpha}(z)=\sum_{\nu=0}^{\infty}\frac{(-1)^{\nu}}{\nu!\,\Gamma(\nu+\alpha+1)}\,\left(\frac{z}{2}\right)^{\alpha+2\nu}\;.
$$
for the Bessel function, and
$$
Ai(y)=\frac{\sqrt[3]{3}}{\pi}\,A(-3\sqrt{3}y),\quad A(t)=\frac{\pi}{3}\,\sqrt{\frac{t}{3}}\, \left[J_{-1/3}\left(2\left(\frac{t}{3}\right)^{\frac{3}{2}}\right)+J_{1/3}\left(2\left(\frac{t}{3}\right)^{\frac{3}{2}}\right)\right].
$$
for the Airy function (see \cite{szego_75}). When the transition regimes dominate in integral (\ref{eq:c1.2}), then the asymptotics of $N_n(\alpha, p, \beta)$ besides the degree on $n$ have the factor $\ln n$. It is also curious to mention that if these regimes dominate, then the gamma factors in the constant $C(\beta,p)$ for the oscillatory cosine regime explode. For the cosine-Bessel regime it happens for $\beta+1-p/2=0$, and for the cosine-Airy regime it happens for $1-p/2=0$.

\newcommand\mc[1]{\multicolumn{1}{c}{#1}}
\begin{table}[H]
\centering
\begin{threeparttable}
    \caption{Asymptotic regimes\textdagger  of the Laguerre norms $N_{n,l}(\alpha,p,\beta)$}
    \label{table1}
\begin{tabular}{ *2l }    \hline
\emph{Asymptotic regime} 
& \emph{Constant}   \\ \hline
Bessel regime    & $C_{B}(\alpha,p,\beta):=2\int\limits_{0}^{\infty}t^{2\beta+1}|J_{\alpha}|^{2p}(2t)\,dt$    \\[3ex] 
Airy regime & $C_{A}(p):=\int_{-\infty}^{+\infty}\left[\frac{2\pi}{\sqrt[3]{2}}\,\, {\rm Ai}^2\left(-\frac{t \sqrt[3]{2}}{2}\right) \right]^p \,dt$\\[3ex]
Cosine regime  & $C(p,\beta):=\displaystyle\frac{2^{\beta+1}}{\pi^{p+1/2}}\,\displaystyle\frac{\Gamma(\beta+1-p/2)\,\Gamma(1-p/2)\,\Gamma(p+1/2)}{\Gamma(\beta+2-p)\,\Gamma(1+p)}$  \\[1.5ex] \bottomrule
 \hline
\end{tabular}
 \begin{tablenotes}
    \item[\textdagger] There also exist two asymptotic transition regimes: cosine-Bessel and cosine-Airy; when they dominate, the asymptotics of $N_{n}(\alpha, p, \beta)$ has a factor $\ln n$ besides the $n$th-power law.
    \end{tablenotes}
  \end{threeparttable}
  \end{table}

The application of this methodology \cite{aptekarev_2015} to the three-dimensional hydrogenic system has allowed us to find the asymptotics ($n\rightarrow \infty$) of the hydrogenic Laguerre norms $N_{n,l}(\alpha, p, \beta)$ given by (\ref{1.2}) and (\ref{eq:c1.3}), and thus the dominant term of $R_p[\rho_{n,l}]$ given by (\ref{eq:renyi_lp}). We have obtained the following values for the radial Rényi entropy $R_p[\rho_{n,l}]$ of the Rydberg hydrogenic states for all possible values of $p$:\\

\begin{equation}
\label{eq:betapos}
R_p[\rho_{n,l}]= \frac{1}{1-p} \ln \Bigg[\frac{n^{3-4p}}{2^{3-2p}Z^{3(1-p)}} \times
\left\{\begin{array}{cc}
 C(p,\beta)\,(2(n-l-1))^{3-2p}\,(1+\bar{\bar{o}}(1))\Bigg], & p\in(0,2)  \\[2ex]
  \hspace{-3cm}  \frac{\ln (n-l-1)+\underline{\underline{O}}(1)}{\pi^{2}(n-l-1)}\Bigg], & p=2 \\[2ex]
 C_{B}(\alpha,p,\beta)\,(n-l-1)^{-(3-p)}\,(1+\bar{\bar{o}}(1))\Bigg], & p\in(2,\infty)\\
  \end{array}\right.
\end{equation}
Note that the Airy regime does not play a significant role at first order in our hydrogenic system. The reason is that for $p=2$ the transition cosine-Bessel regime determines the asymptotics of $N_{n,l}(p=2)$.
Thus, we have (a) for $p\in(0,2)$ the region of $\mathbb{R}_{+}$ where the Laguerre polynomials exhibit the cosine asymptotics contributes with the dominant part in the integral (\ref{eq:c1.2}), and (b) for $p>2$ the Bessel regime plays the main role.\\

Finally, from Eqs. (\ref{eq:betapos}) and taking into account the values $\alpha=2l+1,\,l=0,1,\ldots, n-1$, and $\beta = 2-p$ of the involved parameters, one has the following asymptotics for the radial Rényi entropies of the Rydberg states with the orbital quantum number $l<<n$ (which are the most experimentally accesible ones \cite{lundee}):
\begin{equation}
\label{eq:betapos2}
R_p[\rho_{n,l}]= \frac{1}{1-p} \ln  
\left\{\begin{array}{cc}
C(p)\frac{n^{6(1-p)}}{Z^{3(1-p)}}\,(1+\bar{\bar{o}}(1)),	& p\in(0,2)  \\[2ex]
\frac{ n^{2 - 4 p}}{ 2^{3 - 2 p}Z^{3(1 -p)}}\frac{\ln n + \underline{\underline{O}}(1)}{\pi^{2}}, & p=2 \\[2ex]
C_{B}(l,p)\,\frac{n^{-3p}}{2^{3-2p}Z^{3(1-p)}}\,(1+\bar{\bar{o}}(1)), & p\in(2,\infty)\\
						  \end{array}\right.
\end{equation}
where $C(p) \equiv C(p,\beta=2-p)$ and $C_{B}(l,p) \equiv C_{B}(\alpha=2l+1,p,\beta=2-p) $.

\section{Results and numerical discussion}\label{sec:4}

In this section we obtain the Rényi, Shannon and Tsallis entropies of the Rydberg hydrogenic states in terms of the quantum numbers and the nuclear charge $Z$. Then, for illustration, we numerically discuss the Rényi entropy  $R_{p}[\rho_{n,0,0}]$ of some  Rydberg hydrogenic states $n$\textit{s} in terms of $n$, $p$ and $Z$.\\

First, by putting in Eq. (\ref{eq:totalentro}) the values of the radial Rényi entropy $R_p[\rho_{n,l}]$ given by Eq. (\ref{eq:betapos}) and taking into account the angular Rényi entropy $R_{p}[Y_{l,m}]$ given by Eqs. (\ref{eq:angpart}) and (\ref{eq:angentro}), one obtains the total Rényi entropy $R_{p}[\rho_{n,l,m}]$ of the Rydberg states in a straightforward manner. Second, from the latter expression and Eq. (\ref{eq:rel2}) one can readily obtain the Tsallis entropy $T_{p}[\rho_{n,l,m}]$ of the Rydberg states.\\

Third, a most important case in the previous expressions is the limit $p\rightarrow1$ since then the Rényi entropy $R_{p}[\rho]$ of a probability density $\rho$ is equal to the Shannon entropy $S[\rho]$, as already mentioned above. By keeping in mind Eq. (\ref{eq:totalentro}), to investigate this limiting case for the Rényi entropy $R_{p}[\rho_{n,l,m}]$ of the Rydberg hydrogenic states we first take into account that
\begin{eqnarray}
\label{eq:radshan}
\lim_{p \to +1} R_p[\rho_{n,l}] &=& \lim_{p \to +1} \frac{1}{1-p} \ln \left[\frac{n^{3-4p}}{2^{3-2p}Z^{3(1-p)}} C(p,\beta)\,(2n)^{1+\beta-p}\right]\nonumber\\
&=& 6\ln n - \ln 2 +\ln\pi -3\ln Z ,
\end{eqnarray} 
(where we used (\ref{eq:betapos}) and $l<<n$ at the first equality),  and 
\begin{equation}
\label{eq:angshan}
\lim_{p \to +1} R_{p}[Y_{l,m}] = \lim_{p \to +1} \frac{1}{1-p} \ln \Omega_{l,m}(\theta,\phi) = S[Y_{l,m}],
\end{equation}
(remember (\ref{eq:angentro}) for the first equality) where $S[Y_{l,m}]$ is the Shannon-entropy functional of the spherical harmonics given \cite{dehesa2,dehesa_2010} by 
\begin{equation}
S[Y_{l,m}] = \int_{0}^{\pi} \int_{0}^{2\pi}[Y_{l,m}(\theta,\phi)]^{2}\, \ln \,[Y_{l,m}(\theta,\phi)]^{2}\,\sin\theta\, d\theta d\phi,
\end{equation}
which is under control. Then the limit $p\rightarrow1$ in Eq. (\ref{eq:totalentro}) gives rise, keeping in mind Eq. (\ref{eq:betapos}), to the following value 
\begin{equation}
\label{eq:totalshan}
S[\rho_{n,l,m}] = \lim_{p \to +1} R_p[\rho_{n,l,m}] = 6\ln n - \ln 2 +\ln\pi -3\ln Z + S[Y_{l,m}] + o(1)
\end{equation}
for the Shannon entropy of the Rydberg hydrogenic states. This expression has been previously obtained \cite{lopez_2013} by a different technique, what is a further checking of our results.\\

In particular, from  the previous results we find that the following values
\begin{equation}
\label{eq:nsrenyi}
R_{p}[\rho_{n,0,0}] = R_p[\rho_{n,0}] + R_{p}[Y_{0,0}] = R_p[\rho_{n,0}] + \ln (4\pi)
\end{equation}
for $p \neq 1$, and
\begin{equation}
S[\rho_{n,0,0}] = 6\ln n + \ln 2 + 2\ln\pi -3\ln Z + o(1)	
\end{equation}
for the Rényi and Shannon entropy of the ($n$\textit{s})-Rydberg hydrogenic states, respectively. Here we have used that $S[Y_{0,0}] = \ln (4\pi)$ and the explicit values of $R_p[\rho_{n,0}]$ are given in Eq. (\ref{eq:betapos}).\\

\begin{figure}[H]
\centering
 \includegraphics[width=\linewidth]{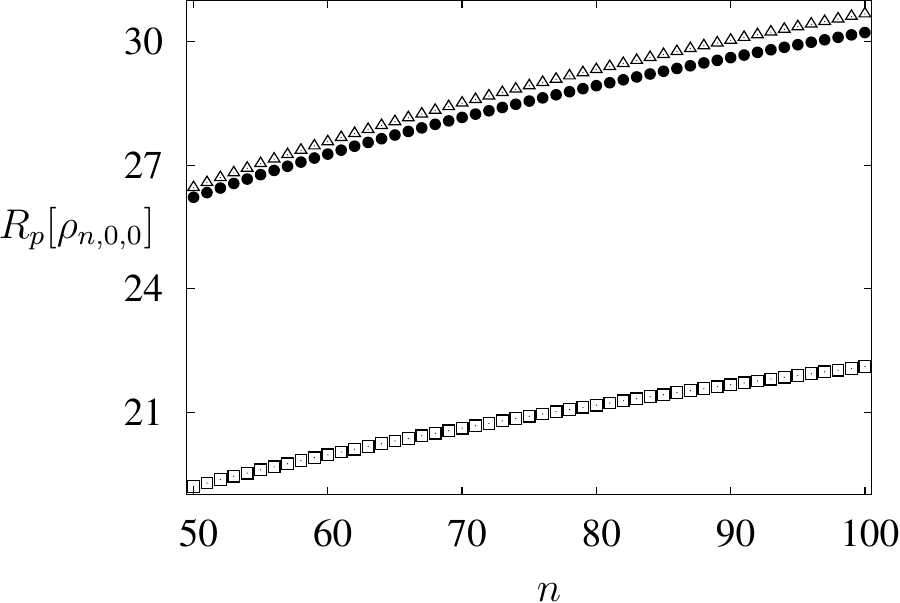}
 \caption{Variation of the Rényi entropy, $R_{p}[\rho_{n,0,0}]$ for the Rydberg $(ns)$-states of the hydrogen atom ($Z=1$) with respect to $n$, when $p=\frac{3}{4}(\triangle), 2(\bullet)$ and $\frac{7}{2}(\square)$.}
 \label{fig:n}
 \end{figure}

Finally, for illustration we numerically study the variation of the Rényi entropy $R_{p}[\rho_{n,0,0}]$ for some Rydberg ($n$\textit{s})-states on the quantum number $n$, the order parameter $p$ and the nuclear charge $Z$. Let us start with the variation of the $p$-th order Rényi entropy of these states in terms of the principal quantum number $n$ when $p$ is fixed. As an example, this quantity with $p=\frac{3}{4}(\triangle), 2(\bullet), \frac{7}{2}(\square)$ is plotted in Fig. \ref{fig:n}. We observe that the behavior of the Rényi entropy of the Rydberg ($n$\textit{s})-states has an increasing character, which can be explained by the fact that the system tends to the classical regime as $n$ increases.\\

\begin{figure}[H]
\centering
 \includegraphics[width=\linewidth]{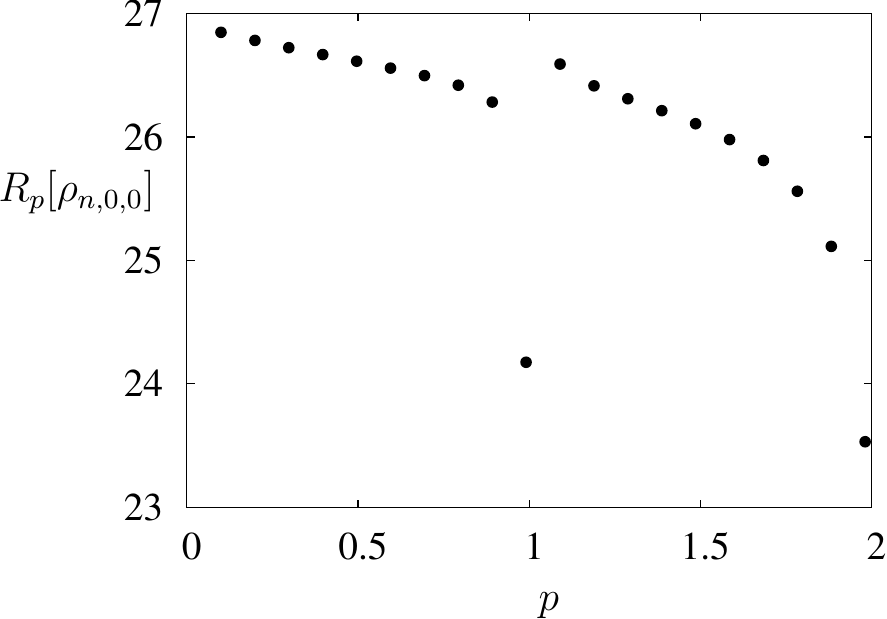}
 \caption{Variation of the Rényi entropy, $R_{p}[\rho_{n,0,0}]$, with respect to $p$ for the Rydberg state with $n=50$ of the hydrogen atom ($Z=1$) when $p\in (0,2)$.}
 \label{fig:p1}
 \end{figure}

\begin{figure}[H]
\centering
 \includegraphics[width=\linewidth]{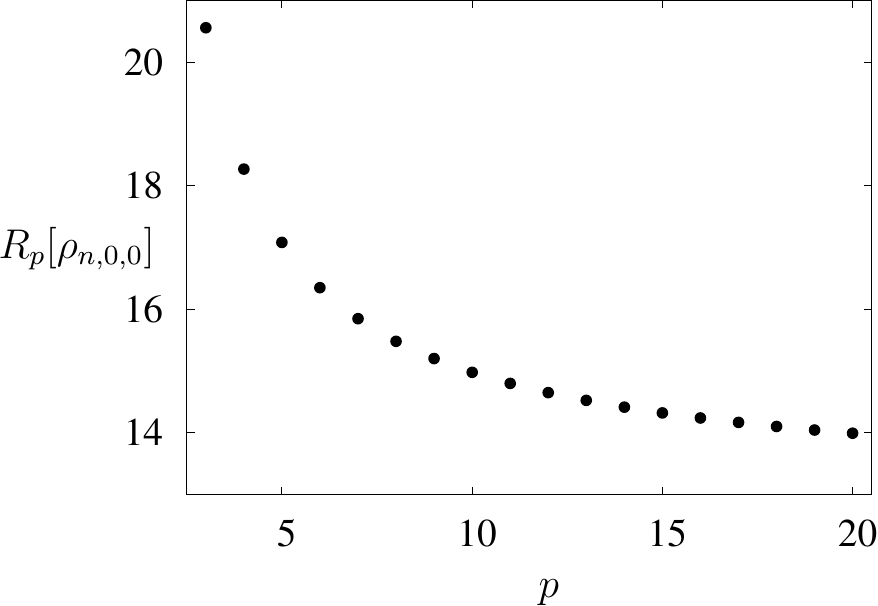}
 \caption{Variation of the Rényi entropy, $R_{p}[\rho_{n,0,0}]$, with respect to $p$ for the Rydberg state with $n=50$ of the hydrogen atom ($Z=1$) when  the integer $p\in (3,20)$.}
 \label{fig:p2}
 \end{figure}

Then, we study in Figs. (\ref{fig:p1})-(\ref{fig:p2}) the variation of the Rényi entropy, $R_{p}[\rho_{n,0,0}]$, with respect to the order $p$, with $p\in(0,20)$, for the Rydberg hydrogenic state which corresponds to $n=50$. Therein we observe that the Rényi entropy decreases as the order $p$ increases. This behavior (a) is not monotonic when $p\in(0,2]$ because the decreasing at $p=1$ and $p=2$ is specially pronounced, and (b) is monotonic when $p>2$. The monotonicity of the latter case is a consequence of the Bessel asymptotic regime.  Moreover, by globally looking at the entropy values with integer $p$ we observe that the entropic quantities with the lowest orders (particularly when $p=1$ and $p=2$, closely related with the Shannon entropy and the desiquilibrium, respectively, as already mentioned) are most significant for the quantification of the spreading of the electron distribution of the system. \\

\begin{figure}[H]
      \centering
          \includegraphics[width=\textwidth]{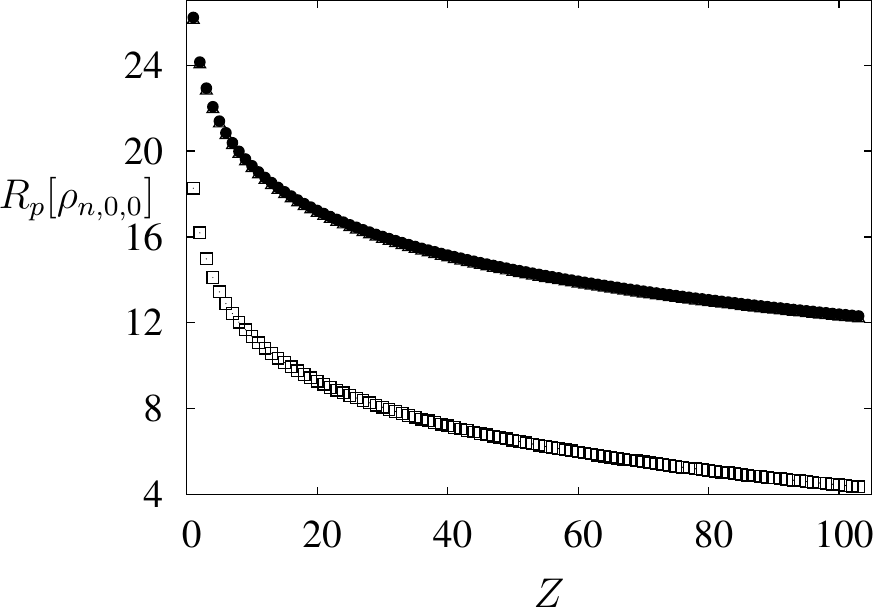}
      \caption{Variation of the Rényi entropy, $R_{p}[\rho_{n,0,0}]$, with respect to the atomic number $Z$ for the Rydberg hydrogenic states with $n=50$ when $p= \frac{3}{2}(\triangle),2(\bullet)$ and $4(\bigodot)$.}
      \label{fig:Z}
    \end{figure}

Finally, in Fig. \ref{fig:Z}, we study the behavior of the Rényi entropy, $R_{p}[\rho_{n,0,0}]$, as a function of the atomic number $Z$ of the Rydberg hydrogenic state with $n=50$ for different values of the order parameter ($p= \frac{3}{2}(\triangle),2(\bullet)$ and $4(\bigodot)$) when $Z$ is ranging from hydrogen $(Z=1)$ to lawrencium $(Z=103)$. We observe in all cases that the Rényi entropy decreases monotonically as $Z$ increases. In particular tha behavior of $R_{2}[\rho_{n,0,0}]$ (whose exponential gives the disequilibrium) points out the fact that the probability distribution of the system tends to separate from equiprobability more and more as the electron number of the atom increases; so, it nicely quantifies the complexity of the system as the atomic number grows. 

\section{Conclusions}\label{sec:5}

In this work we have explicitly calculated the dominant term of the Rényi, Shannon and Tsallis entropies for all quantum-mechanically allowed Rydberg (i.e., highly excited) hydrogenic states in terms of the nuclear charge $Z$ and the quantum numbers which characterize the corresponding state's wavefunctions. We have used a novel technique based on some ideas extracted from the modern approximation theory, which allows us to determine the asymptotics $(n\rightarrow \infty)$ of the $\mathcal{L}_{p}$-norm, $N_{n,l}(p)$, of the Laguerre polynomials which control the associated wavefunctions. Finally, for illustration, we have studied the behavior of the Rényi entropy for the Rydberg ($n$\textit{s})-states at various values of the involved parameters $(n,p,Z)$. We have found that this quantity (a) decreases as a function of $p$, indicating that the most relevant Rényi quantities of integer order are those associated with the Shannon entropy and the disequilibrium, (b) has an increasing character for all Rydberg values of $n$ as the parameter $p$ is increasing, which can be explained by the fact that the system tends to the classical regime as $n$ increases, and (c) decreases for all $p$'s as the nuclear charge is increasing when $n$ is fixed.

\section*{Acknowledgments}
We thank A.I. Aptekarev for useful discussions. This work has been partially supported by the Projects
FQM-7276 and FQM-207 of the Junta de Andaluc\'ia and the MINECO-FEDER grants
FIS2011-24540, FIS2014- 54497 and FIS2014-59311-P. I. V. Toranzo acknowledges the support of ME under the program FPU.

\end{document}